\documentclass[sigconf]{acmart}

\usepackage{booktabs} 
\usepackage{wrapfig}
\usepackage[algoruled,linesnumbered]{algorithm2e}
\usepackage{multirow}
\usepackage{subcaption}

\newcommand{\argmin}{\mathop{\mathrm{argmin}}}
\newcommand{\vv}{\mathbf v}

\newcommand{\bld}{\textbf}
\newcommand{\blu}[1]{\underline{\textbf{#1}}}
\newcommand{\mymodel}{\texttt{BB2vec}}

\makeatletter
\renewcommand\@formatdoi[1]{\ignorespaces}
\makeatother
\setcopyright{rightsretained}
\acmDOI{ }
\acmISBN{}
\setcopyright{rightsretained}
\acmConference[RecSysKTL'18]{Workshop on Intelligent Recommender Systems by Knowledge Transfer and Learning}{October 2018} {Vancouver, Canada}
\acmYear{2018}
\copyrightyear{2018}
\acmArticle{}
\acmPrice{}

\begin{document}
\title{Inferring Complementary Products from Baskets and Browsing Sessions}

\author{Ilya Trofimov}
\affiliation{%
  \institution{Yandex Market}
  \city{Moscow}
  \state{Russia}
}
\email{trofim@yandex-team.ru}

\renewcommand{\shortauthors}{I. Trofimov}

\begin{abstract}
Complementary products recommendation is an important problem in e-commerce.
Such recommendations increase the average order price and the number of products in baskets.
Complementary products are typically inferred from basket data.
In this study, we propose the \mymodel{} model.
The \mymodel{} model learns vector representations of products by analyzing jointly two types of data - \textbf{B}askets and \textbf{B}rowsing sessions (visiting web pages of products).
These vector representations are used for making complementary products recommendation.
The proposed model alleviates the cold start problem by delivering better recommendations for products having few or no purchases.
We show that the \mymodel{} model has better performance than other models which use only basket data.
\end{abstract}

%
%


\keywords{embeddings learning, multi-task learning, recommender systems, e-commerce}
\maketitle

\section{Introduction}

Recommender systems in e-commerce are the ``must have'' technology now.
These systems are used by almost all major e-commerce companies.
Recommender systems help users to navigate in a vast assortment, discover new items and satisfy various tastes and needs.
For online shops, such systems help to convert browsers into buyers (increase conversion rate), do cross-selling,  improve users loyalty and retention \cite{schafer1999recommender}.
These effects overall increase the revenue of an e-commerce company and improve customer experience.
Worldwide retail e-commerce sales in 2016 are estimated as \mbox{\$1,915} Trillion and will continue to grow by 18\%-25\% each year
\footnote{https://www.emarketer.com/Article/Worldwide-Retail-Ecommerce-Sales-Will-Reach-1915-Trillion-This-Year/1014369}. The share of e-commerce in all retail in 2016 is estimated as 8.7\% and continues to grow \footnote{https://www.statista.com/statistics/534123/e-commerce-share-of-retail-sales-worldwide/}.

The main types of recommendations at e-commerce web site are:
\begin{itemize}
\item \textbf{Personalized}: ``Products you may like''.
\item \textbf{Non-personalized}: ``Similar products'', ``Complementary products'', ''Popular products''.
\end{itemize}

Personalized recommendations enjoy a great body of research in recent years \cite{koren2009matrix, Rendle2009, hu2008collaborative, deshpande2004item, Linden2003, wang2011collaborative}.
In the same time, non-personalized recommendations are less studied.
``Similar products'' recommendations are typically placed on the product web page.
Similar products are \textit{substitutes} and can be purchased interchangeably.
The major goal of similar products recommendation is to persuade a user to purchase by presenting him/her a diverse set of products similar to the original interest.
\textit{Complementary} products can be purchased in addition to each other.
For example, an iPhone cover is complementary for iPhone, the second part of a film is complementary to the first part, etc.

When a user already has an intention to purchase - it's high time for complementary products recommendation.
Such recommendations might appear on the product web page, during addition to the shopping cart and the checkout process.
In the latter case, recommendations can be based on the whole content of the basket.
People naturally tend to buy products in bundles.
Complementary products recommendation satisfies this natural need and increases average order price as a consequence.

Online shops try to maximize the revenue by combining all the recommendation scenarios.
The significance of recommendations in e-commerce can be illustrated by the following fact: in Amazon, 35\% of purchases overall come from products recommendations \footnote{http://www.mckinsey.com/industries/retail/our-insights/how-retailers-can-keep-up-with-consumers}.

These considerations show that complementary products recommendation is an important scenario.
Manual selection of complementary products fails when the number of products in an online shop (or a marketplace) is large and assortment changes quickly.
Some companies \footnote{For example \url{richrelevance.com}, \url{gravity.com}, \url{yoochoose.com}, etc.} provide ``recommendations as a service'' for online shops.
These companies deliver real-time recommendations for hundreds and even thousands of shops from various business areas in a fully automated manner.

Machine learning models can solve this problem.
There are two principled ways for making complementary products recommendations via machine learning.
First way - is to use human assessment to generate the ``ground truth'' - a set of product-complement pairs.
Then some machine learning model can use these pairs as positive examples and build a classifier.
This classifier could be applied for identifying complements for all products.
The application of this approach is limited for two main reasons. Firstly, it relies on the ``ground truth'' which should be collected by human assessment.
Human assessment is costly and error-prone when the number of products is large and the assortment changes quickly.

An alternative approach is inferring complementary products by analyzing user purchases (baskets) and detecting items which are \textit{frequently purchased together}.
To the best of our knowledge, it is typically done in e-commerce companies by using some heuristical co-occurrence measure (cosine similarity, Jaccard similarity, PMI).
The more sophisticated way is to develop a predictive model for such co-occurrences in baskets. In this study, we follow the latter approach.

We make the following contributions in this paper:
\begin{enumerate}
\item We propose the statistical model \mymodel{}. The model itself learns vector representations of products by analyzing jointly two types of data - baskets and browsing sessions (visiting web pages of products).
These types of data are always available to any e-commerce company.
We apply vector representations learned by the \mymodel{} model for doing recommendations of complementary products.
\item In experimental studies, we prove that the proposed model \mymodel{} predicts products which are purchased together better than other methods which rely only on basket data.
We show that the \mymodel{} model alleviates the cold-start problem by delivering better predictions for products having few or no purchases.
\item We show how to make the \mymodel model scalable by selecting the specific objective function. It is an important issue since e-commerce company typically has a vast amount of browsing data, much more than basket data.
\end{enumerate}

The source code of our model is publicly available \footnote{\texttt{https://github.com/IlyaTrofimov/bb2vec}}.

\section{Related work}

After invention of the \texttt{word2vec} model \cite{Mikolov2013a, Mikolov2013b} which is dedicated to learning of distributed word representations, it was rapidly extended to other kinds of data: sentences and documents
\cite{Le2014} (\texttt{doc2vec}), products in baskets \cite{Grbovic2015} (\texttt{prod2vec}, \texttt{bagged-prod2vec}),
nodes in graph \cite{grover2016node2vec} (\texttt{node2vec}), \cite{perozzi2014deepwalk} (\texttt{DeepWalk}), etc.
The original \texttt{word2vec} algorithm in its skip-gram variant proved to solve the weighted matrix factorization problem, where the matrix consists of values of shifted PMI of words \cite{Levy2014}.
The extension of the \texttt{prod2vec} for using items metadata was proposed in \cite{Vasile2016} (\texttt{MetaProd2Vec}) and for the textual data in \cite{djuric2015hierarchical}.
Sequences of items selected by a user are modeled in \citep{guardia2015latent} by means of the \texttt{word2vec}-like model.
User and item representations learned by their model are applied for making recommendations.

Another associated area of research is \textit{multi-relational learning}.
Multi-relational data is a kind of data when entities (users, items, categories, time moments, etc.) are connected to each other by multiple types of relations (ratings, views, etc.)
Collective matrix factorization \citep{singh2008relational} was proposed for relational learning.
Each matrix corresponds to a relation between items. Latent vectors of entities are shared.
Other variants of models for multi-relational learning are: \texttt{RESCAL} \cite{nickel2011three}, \texttt{TransE} \cite{Bordes2013}, \texttt{BigARTM} \cite{vorontsov2015non}.

In computational linguistics, distributed word representations for two languages could be learned jointly via
multi-task learning \cite{klementiev2012inducing}. This approach improves machine translation.

The importance of learning-to-rank in context of recommendations with implicit feedback is discussed in \cite{Rendle2009}.

\subsection{The most similar studies}

\textbf{Sceptre} \citep{mcauley2015inferring} consider two types of relationships between products ``being substitutable'' and ``being complementary''.
The ``Sceptre'' model builds two directed graphs for inferring these two types of relations.
As a ground truth for training the model uses recommendation lists crawled from \textit{\url{amazon.com}}.
The topic model based on user reviews which also takes into account items taxonomy was used for feature generation.
At the top level, the logistic regression did the classification.
The application of this approach is limited for two main reasons. Firstly, it relies on ``ground truth'' which in the real situation should be collected by human assessment.
Human assessment is costly and error-prone when the number of products is large.
Secondly, it requires rich text descriptions (product reviews) for doing topic modeling.

The \textbf{prod2vec} \citep{Grbovic2015} model learns product representations from basket data.
These representations are used further for doing recommendations of type ``Customers who bought $X$ from vendor $V_1$ also bought $Y$ from vendor $V_2$''
which are shown alongside emails in a web interface.
Our model is a combination of several \texttt{prod2vec} models which are learned simultaneously with partially shared parameters.
We show that this combination has better predictive performance than the single \texttt{prod2vec} model.

\textbf{P-EMB}. 
Modeling embeddings for exponential family distributions were presented in \cite{Rudolph2016}.
The particular model of this family is ``Poisson embeddings" (\texttt{P-EMB}) which was applied to model quantities of products in market baskets.
Vector representations of products learned by the \texttt{P-EMB} model can be used for inferring substitutable and complementary products.
The application of exponential distributions is orthogonal to our research. We consider the \mymodel{} model can be modified accordingly, i.e.,
learn several \texttt{P-EMB} models with partially shared parameters instead of \texttt{prod2vec} models.

\textbf{CoFactor} \citep{Liang2016} improves recommendations from implicit feedback via weighted matrix factorization.
The improvement is done by the regularization term which is a factorization problem of a matrix with items co-occurrences (shifted PMI) which is also used in our paper.
Our research is different in three main points.
Firstly, two parts of the objective (\ref{obj-purchased-together}) come from different sources of data (baskets and browsing sessions),
while two parts of the objective in \citep{Liang2016} (weighted matrix factorization term and regularization term) come from the one kind of data - user-item implicit feedback.
Secondly, the objective of the \mymodel{} is mixed: it includes the skip-gram objective and the matrix factorization one.
The procedure of inference is also different - SGD, while \citep{Liang2016} used coordinate descent.
Thirdly, we use pairwise ranking objective and show its benefits.

\section{Learning of word representations}

The popular \texttt{word2vec} model \cite{Mikolov2013a, Mikolov2013b} in skip-gram variant
learns vector representations of words which are useful for predicting the context of the word
in a document. More formally, let $\{w_1, \ldots, w_T\}$ be a sequence of words, $W$ - their vocabulary,
the context - a c-sized window of words surrounding $w_t$. The likelihood of such model is
\begin{equation}
\label{word2vec-window}
\frac{1}{T} \sum_{t=1}^{T} \sum_{-c \le j \le c, \, j \neq 0} \log P(w_{t+j} | w_t)
\end{equation}


The conditional distribution $P(w_O | w_I)$ is defined as
\begin{equation}
\label{cond-distr}
P(w_O | w_I) = \frac{\exp{(\vv'_{w_O}}^T \vv_{w_I})}{\sum_{w \in W} \exp({\vv'_{w}}^T \vv_{w_I})}
\end{equation}

where $\vv_w, \vv'_w \in \mathbb{R}^d$ are input and output representations of words. Together they form $|W| \times d$ matrices $V, V'$.
Vectors $\vv_w, \vv'_w$ are also known as \textit{word embeddings} and have many applications in natural language processing, information retrieval, image captioning, etc.
Words of similar meaning typically have similar vector representations compared by cosine similarity.

The exact minimization of (\ref{word2vec-window}) is computationally hard because of the softmax term.
The \texttt{word2vec} algorithm is only concerned with learning high-quality vector representations and it minimizes the more simple \textit{negative sampling} objective

\begin{gather}
\label{neg-problem}
Q^{SG}(V', V) = \frac{1}{T} \sum_{t=1}^{T} \sum_{-c \le j \le c, \, j \neq 0} L_{neg}(w_t, w_{t + j})\\
L_{neg}(w_I, w_O) = \log \sigma({\vv'_{w_O}}^T \vv_{w_I}) + \sum_{i = 1}^k \mathbb{E}_{w_i \sim N(w)} \log \sigma(-{\vv'_{w_i}}^T \vv_{w_I}) \notag
\end{gather}
where superscript $SG$ stands for ``skip-gram''.

In the negative sampling objective, positive examples $(w_I, w_O)$ come from the training data, while in negative samples $(w_I, w_i)$
output words $w_i$ come from a noise distribution $N(w)$.
The noise distribution $N(w)$ is a free parameter which in the original \texttt{word2vec} algorithm is the unigram distribution to the 3/4rd power - $P^{3/4}(w)$.

The \texttt{word2vec} model in skip-gram formulation when $N(w) = P(w)$ implicitly factorizes the matrix of shifted pointwise mutual information (PMI) \cite{Levy2014}:
\begin{equation}
\label{word2vec-nature}
{\vv'_{w_i}}^T \vv_{w_j} = PMI(w_i, w_j) - \log k
\end{equation}
\begin{equation}
\label{spmi}
PMI(w_i, w_j) = \log\left(\frac{n_{ij} / T}{(n_i / T) (n_j / T)}\right)
\end{equation}
where $n_i$ is a number of times which a word $w_i$ occurred in a document, $n_{ij}$ - number of times which two words $w_i$, $w_j$ occurred together in a c-sized window,
$T$ is length of a document.
Let
$$
Q^{MF}(V', V) = \frac{1}{2} \sum_{i,j} (PMI(w_i,w_j) - \log k - {\vv'_{w_i}}^T \vv_{w_j})^2
$$
where superscript $MF$ stands for ``matrix factorization''.
Solutions of these two problems
\begin{equation}
\label{sg-mf-sim}
\argmin_{V', V} Q^{SG}(V', V) \qquad \argmin_{V', V} Q^{MF}(V', V)
\end{equation}
are equal for large enough dimensionality $d$ of vector representations \cite{Levy2014}.

\section{The proposed model}
In this section, we describe the proposed \mymodel{} model.
It is a combination of \texttt{prod2vec} models which are fitted with partially shared parameters (multi-task learning).

\subsection{Product recommendations with \texttt{prod2vec}}
\label{sec-prod2vec}

\begin{figure}
\includegraphics[width=0.5\textwidth]{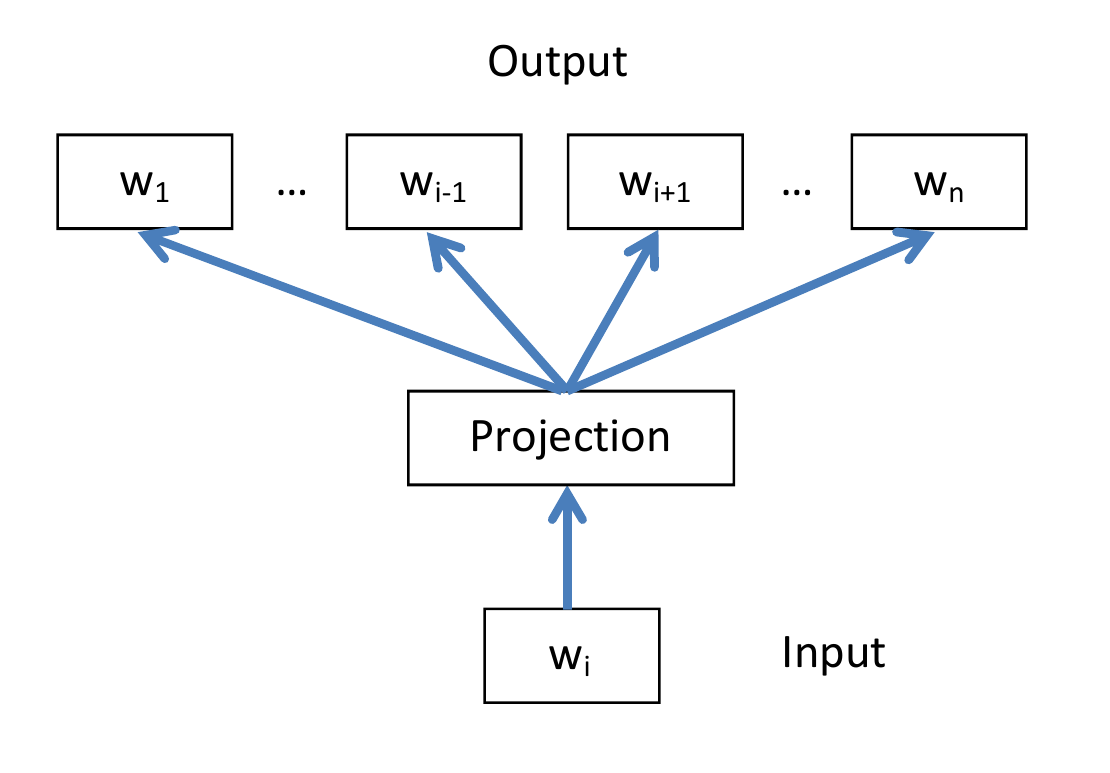}
\caption{Skip-gram model, basket/session = $\{w_1, w_2, \ldots, w_n \}$}
\label{fig:skip-gram}
\end{figure}

The original \texttt{word2vec} model can be easily applied to other domains by redefining the ``context'' and the ``word''.
Some examples include modeling products in baskets \cite{Grbovic2015}, nodes in graph \cite{grover2016node2vec, perozzi2014deepwalk}, sequences of items selected by users \cite{guardia2015latent}, etc.

Particulary, the \texttt{prod2vec} model \cite{Grbovic2015} assumes the following likelihood
\begin{equation}
\label{prod2vec}
\sum_{B \in \mathfrak{B}} \sum_{w_{t} \in B}\; \sum_{-c \le j \le c, \, j \neq 0} \log P(w_{t+j} | w_t)
\end{equation}
where $\mathfrak{B}$ is a set of baskets.

The context $C$ in the original \texttt{word2vec} model (\ref{word2vec-window}) is a c-sized windows since it is tailored for long text sequences.
In our study, we analyze shorter sequences like baskets or e-commerce browsing sessions.
We use a modified likelihood
\begin{equation}
\label{word2vec-context}
 \sum_{B \in \mathfrak{B}} \sum_{\substack{w_{I}, w_{O} \in B \\ w_{I} \neq w_{O}}} \log P(w_{O} | w_I).
\end{equation}
where context is the full basket, except an input product, independently of its size (see fig. \ref{fig:skip-gram}).
This eliminates the need for selecting the window size.


Consider the product $k$.
By definition, complementary products are frequently purchased together and have high conditional probability $P(m \, | \,k)$
\footnote{Some authors \cite{deshpande2004item} propose the modified expression $\frac{P(m \, | \, k)}{P^\alpha(m)}$ to reduce the bias towards popular items.
In this paper we use $\alpha = 0$, however, the proposed model could by extended to the general case $\alpha > 0$.}.


The most simple way to estimate the conditional probability $P(m \, | \, k)$ is by empirical frequencies (co-counting).
Let $n_k$ be a number of baskets with item $k$, $n_{mk}$ - number of baskets with both items $k$ and $m$.
Here and after we assume for simplicity that an item might appear in a basket/session only once.
Then
$$
P(m \, | \, k ) \approx \frac{n_{mk}}{n_k}
$$
However, counts $n_{mk}, n_k$ are noisy for products having few purchases.

The \texttt{prod2vec} model can generate recommendations of complementary products without co-counting.
Given the vector representations of products $V, V'$, complementary products for the item $k$ could be inferred by selecting items $m$ with a high score ${\vv'_m}^T \vv_k$
since $P(m \, | \, k) \sim \exp({\vv'_m}^T \vv_k)$ (\ref{cond-distr}).
The connection of representations $V, V'$ to the conditional distribution $P(m \, | \, k)$ in the data is an intrinsic limitation of the \texttt{prod2vec} model.
In the next section, we will show how to overcome this limitation by means of the multi-task learning and the specific objective function.

\subsection{Multi-task learning}

\begin{algorithm}[t]

\caption{Multi-task learning of representations}
\label{alg:fit}

\DontPrintSemicolon
\SetAlgoVlined

\SetKwInOut{Input}{Input}\SetKwInOut{Output}{Output}
\SetKwFor{While}{repeat}{}{}%

\Input{Tasks  $T_i$, weights $\lambda_i$,
$i=1, \ldots, K$\;
}
\BlankLine

\For{$j = 1, \ldots, L$} {
Initialize elements of matrices $V_j, V'_j$ with $N(0, 0.1)$ \;
}

$W = \sum_{i=1}^{K} \lambda_i$ \;
$\xi $ - random variable having categorical distribution $Cat(\frac{\lambda_1}{W}, \ldots, \frac{\lambda_K}{W})$

\While {until convergence} {
    Sample $i$ from $\xi$ \,\,\, (select the task $T_i$)\;
    Sample object-context pair $(w_I, w_O)$ from the task $T_i$ \;
    \For{$m = 1, \ldots, {\mid W \mid }, \, n = 1, \ldots, d$} {
    \begin{gather*}
    V_{g(i)}^{m,n} \gets V_{g(i)}^{m,n} - \eta^{m,n} \frac{\partial L_{neg}(w_I, w_O)}{\partial V_{g(i)}^{m,n}} \\
    {V'}_{g'(i)}^{m,n} \gets {V'}_{g'(i)}^{m,n} - {\eta'}^{m,n} \frac{\partial L_{neg}(w_I, w_O)}{\partial {V'}_{g'(i)}^{m,n}}
    \end{gather*}
    where learning rates $\eta^{m,n}, {\eta'}^{m,n}$ are calculated by the AdaGrad rule \cite{duchi2011adaptive}.
    }
}

\BlankLine
\Output{$V_j, V'_j$, for $j = 1 \ldots L$}
\end{algorithm}

Assume that multiple types of data contain same objects. E-commerce data naturally have such data types: browsing sessions, baskets, product comparisons, search results, etc.
Let \{$T_1, \ldots, T_K\}$ be a list of such data types.
For each data type $T_i$, one may fit the \texttt{prod2vec} model and infer latent representations of items $V_i, V'_i$ by solving the problem
$$
\min_{V_i', V_i} Q^{SG}_{T_i}(V'_{i}, V_{i})
$$
We will refer to each learning problem as a \textit{task}.

Then we assume that some representations are shared and the total number of distinct representations $L$ is less then the number of tasks $K$.
The functions $g', g: \{1, \ldots, K\} \to \{1, \ldots, L\}$ define which matrices are shared between tasks.
For example, if $g(i) = g(j)$ then the input representations $V$ of objects in the tasks $T_i, T_j$ are shared.
We come to the following multi-task learning problem
\begin{align}
\label{sel}
& \min_{\substack{V_j', V_j \\ j = 1 \ldots L}} \sum_{i = 1}^{K} \lambda_i Q^{SG}_{T_i}(V_{g'(i)}', V_{g(i)})
\end{align}
Algorithm \ref{alg:fit} describes a procedure for solving the problem (\ref{sel}).
At the top level, the Algorithm \ref{alg:fit} selects a task $T_i$ with probability proportional to its weight $\lambda_i$.
Then a random object-context pair is sampled and representations of items $V_{g(i)}, V'_{g'(i)}$ are updated via SGD with AdaGrad learning rates.

\subsection{Motivation of multi-task representations learning}

Optimizing the objective (\ref{sel}) is an example of \textit{multi-task learning} (MTL). Multi-task learning is about solving several related learning problems
simultaneously. The problems with not enough data may benefit from coupling parameters with other problems.
However, multi-task learning may worsen a predictive performance. Avoiding it requires choosing a proper way of binding tasks together.

Consider a problem of \textit{complementary} items recommendation in e-commerce.
By definition, complementary items are those which could be purchased in addition to each other.
As we explained in the section \ref{sec-prod2vec}, the \texttt{prod2vec} model
learns representations $V_B', V_B$ of products by solving the problem
\begin{equation}
\label{prod2vec}
V'_B, V_B = \argmin_{V', V} Q^{SG}_{baskets}(V', V)
\end{equation}
and products having high score ${\vv'_{m, B}}^T \vv_{k, B}$ are complementary.

This predictive model can be improved further by analyzing another source of data - browsing sessions.
The probability of purchase in a session is typically 1\%-5\%.
Thus, an online shop has much more browsing data then purchasing data.
The \texttt{prod2vec} model can learn product representations from browsing data
by solving the problem
\begin{equation}
\label{prod2vec-sessions}
V'_S, V_S = \argmin_{V', V} Q^{SG}_{brows.}(V', V)
\end{equation}
It is generally accepted that products which are frequently viewed in the same sessions are \textit{similar}.
Let $S$ be a browsing session. Two items $m, k$ having high conditional probability $P(m \in S\, | \, k \in S)$ are similar.

We conclude that conditional distributions $P(m \in B\, | \, k \in B)$ and $P(m \in S\, | \, k \in S)$ are different
which leads to different representations $V'_B, V_B$ vs. $V'_S, V_S$.
In the same time, the general property of the \texttt{prod2vec} and other extensions of the \texttt{word2vec} model is that similar objects have similar representations.

In our research, we use the specific objective function for learning product representations where input $V_B$ and output $V'_B$ representations are shared separately between tasks
\begin{equation}
\label{obj-purchased-together-sg}
\min_{V'_B, V_B, V'_S, V_S} Q^{SG}_{bask.}(V'_B, V_B) + \lambda \left( Q^{SG}_{browse}(V'_S, V_B) + Q^{SG}_{browse}(V'_B, V_S) \right)
\end{equation}
Addition of $\lambda Q^{SG}_{browse}(V'_S, V_B)$ forces $V_B$ to be a good input representation for modeling similar products, particularly
forcing representations of similar products $V_B$ to be close.
The same holds for output representation $V'_B$ because of the term $\lambda Q^{SG}_{browse}(V'_B, V_S)$.

\subsection{Ranking}

Learning of vector representations in skip-gram variant (\ref{neg-problem}) is considered to be a binary classification problem when positive object-context examples are drawn from the training dataset, and negative examples are drawn from the noise distribution.
Since our final goal is to use representations for generating top-N recommendations one may reformulate the problem as learning to rank.
We can easily transform a skip-gram negative sampling problem (\ref{neg-problem}) into the pairwise ranking problem.

In this formulation we treat $w_I$ as a \textit{query item} and force our model to give positive example $w_O$ higher rank than negative examples
$w_i \sim N(w)$. The value ${\vv'_{w_i}}^T \vv_{w_I}$ is the ranking score:
\begin{equation}
\label{learning-to-rank}
Q^{SG}(V', V) = \sum_{ \substack{w_I, w_O \in B \\ w_I \neq w_O} }\; \sum_{i = 1}^k \mathbb{E}_{w_i \sim N(w)} \log(\sigma(({\vv'_{w_O}}^T - {\vv'_{w_i}}^T) \vv_{w_I}))
\end{equation}
We can show (see Appendix \ref{ranking-explained}) that optimizing (\ref{learning-to-rank}) with respect to $V, V'$
in the general case $N(w) = P^{\alpha}(w)$ leads to the solution
$$
(\vv'_r - \vv'_m)^T \vv_k = \log \left( \frac{P(r \,|\, k)}{P^\alpha(r)} \right) - \log \left( \frac{P(m \,|\, k)}{P^\alpha(m)} \right)
$$
Thus, if ${\vv'_r}^T \vv_k > {\vv'_m}^T \vv_k$ then $\frac{P(r \,|\, k)}{P^\alpha(r)} > \frac{P(m \,|\, k)}{P^\alpha(m)}$.
Interestingly, the number of negative samples $k$ vanished and doesn't introduce a shift like in the classification setting (\ref{spmi}).
For ranking, it can be considered as a hyperparameter of training.
We will use further the learning to rank (\ref{learning-to-rank}) variant of negative sampling problem.

\subsection{Improving computational performance of the \mymodel{} model}
\label{sec:comp-perf}

Since there are much more browsing data than basket data, solving the skip-gram problems $Q^{SG}_{browse}(\cdot)$
requires much more SGD updates in the Algorithm \ref{alg:fit} than in the original problem $Q^{SG}_{bask.}(\cdot)$ and significantly slows down the program.

The software implementation of the \mymodel{} model solves matrix factorization problems $Q^{MF}_{browse}(\cdot)$ instead of the skip-gram problems $Q^{SG}_{browse}(\cdot)$ for browsing sessions
since their solutions are asymptotically equal (\ref{sg-mf-sim}).
As a result, the software implementation optimizes the following objective
\begin{equation}
\label{obj-purchased-together}
\min_{V'_B, V_B, V'_S, V_S} Q^{SG}_{bask.}(V' _B, V_B) + \lambda \left( Q^{MF}_{browse}(V'_S, V_B) + Q^{MF}_{browse}(V'_B, V_S) \right)
\end{equation}
The matrix factorization problems are solved by the stochastic matrix factorization \cite{koren2009matrix} with AdaGrad learning rates \cite{duchi2011adaptive}.


The complexity of one epoch for solving $Q^{SG}_{browse}(\cdot)$ is
$$
O(\#sessions \cdot avg. session\,size \cdot \#neg.samples \cdot d)
$$
while the complexity of one epoch of the stochastic matrix factorization of the shifted $PMI$ matrix is
$$
O(\#entries\; in\; SMPI\; matrix \cdot d)
$$
For our datasets, the complexity of solving the stochastic matrix factorization of the shifted $PMI$ matrix is roughly $10^3$ times less.


\section{Experiments}

In computational experiments, we compared different models for complementary products recommendation.
We measured how well each model predicts products which are purchased together in baskets at the hold-out data.

\subsection{Datasets}
\label{sec:datasets}

\begin{table*}[t]
  \caption{Datasets.}
  \label{tbl:datasets}
  \begin{tabular}{cccccccc}
    \toprule
    \multirow{2}{*}{\textbf{Dataset}} & \multirow{2}{*}{\textbf{Sessions}} & \multicolumn{3}{c}{\textbf{Baskets}} & \multicolumn{2}{c}{\textbf{Items}} \\
    \cline{3-7}
            &          & train & val/test & w/ $\ge 2$ items & total & w/o purch. at train\\
    \midrule
    RecSys'15 & 9,249,729 & 346,554 & 74,161 & 49\% & 7226 & 0\% \\
    RecSys'15 - 10\% & 9,249,729 & 34,753 & 74,161 & 49\% & 7226 & 17\% \\
    CIKM'16 & 310,486 & 9380 & 2063 & 19\% & 11244 & 11\% \\
    CIKM'16, Categories & 310,486 & 9380 & 2063 & 17\% & 750 & 5\% \\
   \bottomrule
\end{tabular}
\end{table*}

To evaluate the performance of the \mymodel{} and compare it with baselines we used 4 datasets with user behaviour log on e-commerce web sites (see Table \ref{tbl:datasets}):
\begin{enumerate}
\item \bld{RecSys'15} - the dataset from the ACM RecSys 2015 challenge
\footnote{http://2015.recsyschallenge.com/challenge.html}.
The challenge data has two main parts: clicks on items (equivalent to visiting products web pages) and purchases.
We consider a basket to be a set of items purchased in a session.
Products with less then 10 purchases in the whole dataset were removed.
We randomly split sessions to train, validation and test in the proportion 70\%  /  15\%  / 15\%.
\item \bld{RecSys'15 - 10\%} - is the modification of the \bld{RecSys'15} dataset. The only difference is that the training dataset was replaced by it's 10\% subsample, while validation and test were left unchanged.
\item \bld{CIKM'16} - is the dataset from CIKM Cup 2016 Track 2 \footnote{https://competitions.codalab.org/competitions/11161}.
We took browsing log (product page views) and transactions (baskets).
We randomly split sessions to train, validation and test in proportion 70\%  /  15\%  / 15\%.
All baskets have session identifier and were split accordingly.
\item \bld{CIKM'16, Categories} - is a modifications of \bld{CIKM'16} dataset. Since \bld{CIKM'16} dataset is very sparse, we replaced product identifiers with their category identifiers. The number of distinct categories is much less than the number of products.
Thus, dataset becomes denser.
\end{enumerate}


Finally, for each dataset, we left at test and validation parts only products having at least one purchase or a view at the training dataset.
Otherwise, methods under considerations can't learn vector representations for such products.
All the purchase counts and the view counts were binarized.

In computational experiments, PMI matrices were calculated using browsing data from train datasets only.
For RecSys'15, RecSys'15, 10\% datasets the cells of PMI matrix having $n_{ij} \ge 10$ were kept.
For CIKM'16, CIKM'16 Category the cells of PMI matrix having $n_{ij} \ge 3$ were kept.

\subsection{Evaluation}

We evaluated the performance of the models by predicting items which are purchased together in baskets at hold-out data.
We used two measures: average HitRate@K and NDCG@K.
For each basket $B$, consider all distinct pairs of items $k, m \in B$.
The goal is to predict the second item $m$ in the pair given the first one $k$ (query item).
Denote $L_k$, a list of length K with complementary items recommendations for the query item $k$.

Then
\begin{gather*}
\text{HitRate@K} = [m \in L_k] \\
\text{NDCG@K} = \frac{[m \in L_k]}{\log_2(pos(m, L_k) + 1)}
\end{gather*}
where $[\cdot]$ is an indicator function, $pos(m, L_k)$ is the position of item $m$ in the list $L_k$.
These performance measures are averaged
\begin{itemize}
\item over all distinct pairs of items $(k, m)$ in all the baskets;
\item or over a subset of pairs of items $(k, m)$ with query item $k$ having a particular number of purchases at the training dataset.
\end{itemize}

\subsection{Models}

\begin{table*}[!]
  \center
  \caption{Experimental results.}
  \label{tbl:res}
  \begin{tabular}{ccccccccc}
    \toprule
     \multirow{3}{*}{Method} & \multicolumn{4}{c}{CIKM'16} & \multicolumn{4}{c}{CIKM'16, Categories} \\
    \cline{2-9}
     & \multicolumn{2}{c}{HitRate@\qquad} & \multicolumn{2}{c}{NDCG@\;\;} & \multicolumn{2}{c}{HitRate@\;\;} & \multicolumn{2}{c}{NDCG@\;\;} \\
    \cline{2-9}
     & 10 & 50 & 10 & 50 & 10 & 50 & 10 & 50 \\
    \hline
    Popularity                & 0.006       & 0.047       & 0.006       & 0.024                 & 0.201 & 0.482 & 0.174 & 0.305 \\
    Co-counting               & 0.015       & 0.016       & 0.023       & 0.024                 & 0.391 & 0.477 & 0.531 & 0.574 \\
    \texttt{Prod2Vec}         & 0.018       & 0.021       & 0.025       & 0.026                 & 0.322 & 0.510 & 0.467 & 0.547 \\
    \mymodel{}, class.      & 0.027       & 0.039       & \blu{0.036} & 0.042                 & 0.354 & 0.609 & 0.483 & 0.598  \\
    \mymodel{}, ranking     & \blu{0.029} & \blu{0.077} & 0.029       & \blu{0.051}           & \blu{0.430} & \blu{0.659} & \blu{0.559} & \blu{0.665} \\
    \bottomrule
\end{tabular}

\vskip.1in

  \begin{tabular}{ccccccccc}
    \toprule
    \multirow{3}{*}{Method} & \multicolumn{4}{c}{RecSys'15} & \multicolumn{4}{c}{RecSys'15, 10\%} \\
    \cline{2-9}
     & \multicolumn{2}{c}{HitRate@} & \multicolumn{2}{c}{NDCG@} & \multicolumn{2}{c}{HitRate@} & \multicolumn{2}{c}{NDCG@} \\
    \cline{2-9}
      & 10 & 50 & 10 & 50 & 10 & 50 & 10 & 50 \\
    \hline
    Popularity             & 0.036 & 0.127 & 0.031 & 0.071 &                    0.036 & 0.128 & 0.031 & 0.072 \\
    Co-counting            & \blu{0.383} & 0.569 & \blu{0.475} & 0.561 &        0.333 & 0.453 & 0.419 & 0.475 \\
    \texttt{Prod2Vec}      & 0.379 & 0.585 & 0.461 & 0.557 &                    0.329 & 0.411 & 0.400 & 0.440 \\
    \mymodel{}, class.   & \blu{0.383} & 0.593 & 0.464 & 0.562 &        0.351 & 0.505 & 0.420 & 0.493 \\
    \mymodel{}, ranking  & \blu{0.383} & \blu{0.597} & 0.465 & \blu{0.564} &        \blu{0.356} & \blu{0.559} & \blu{0.425} & \blu{0.519} \\
    \bottomrule
\end{tabular}
\end{table*}

\begin{figure*}
\center
        \begin{subfigure}[tp]{0.49\textwidth}
                \includegraphics[width=\textwidth]{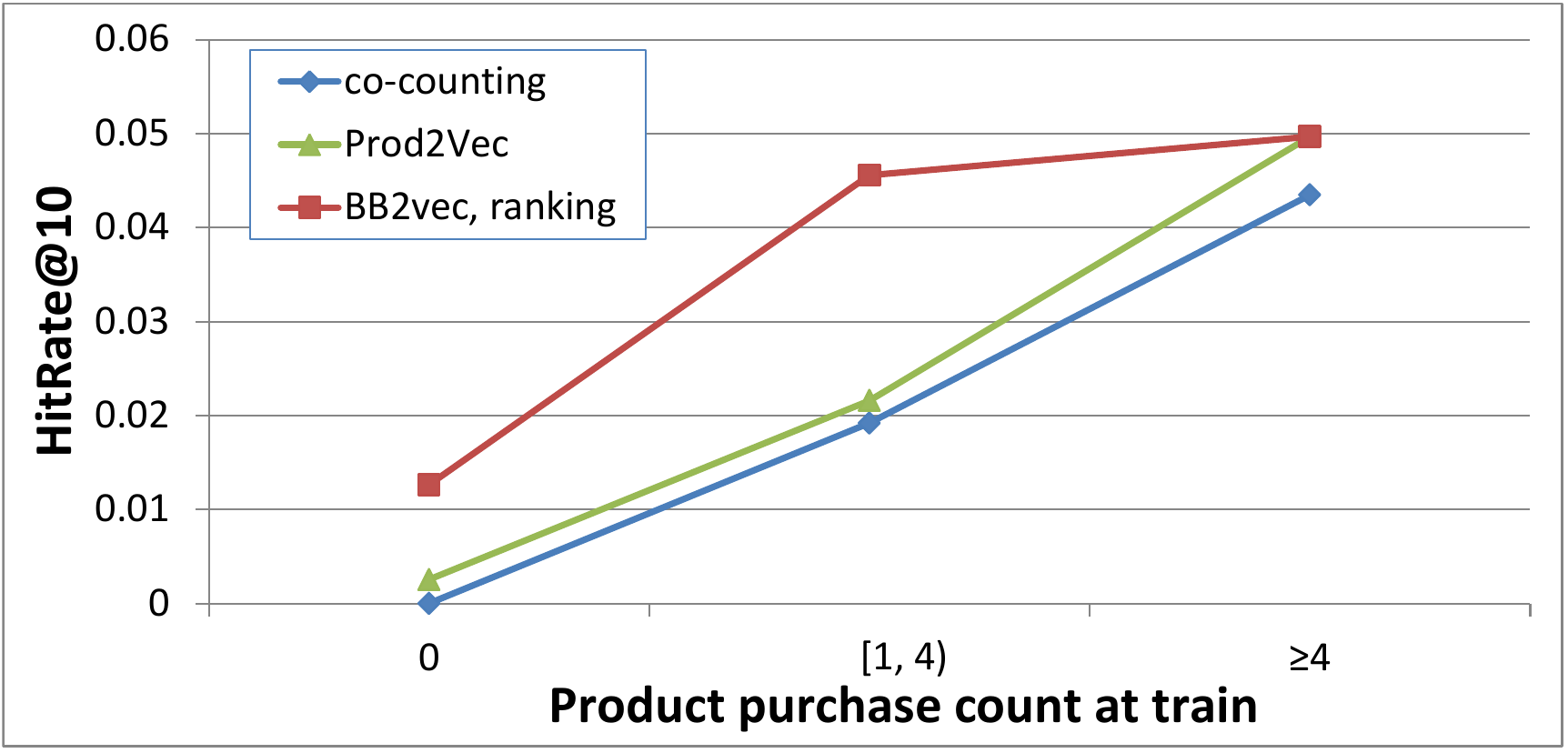}
                \caption{CIKM'16}
        \end{subfigure}
        \begin{subfigure}[tp]{0.49\textwidth}
                \includegraphics[width=\textwidth]{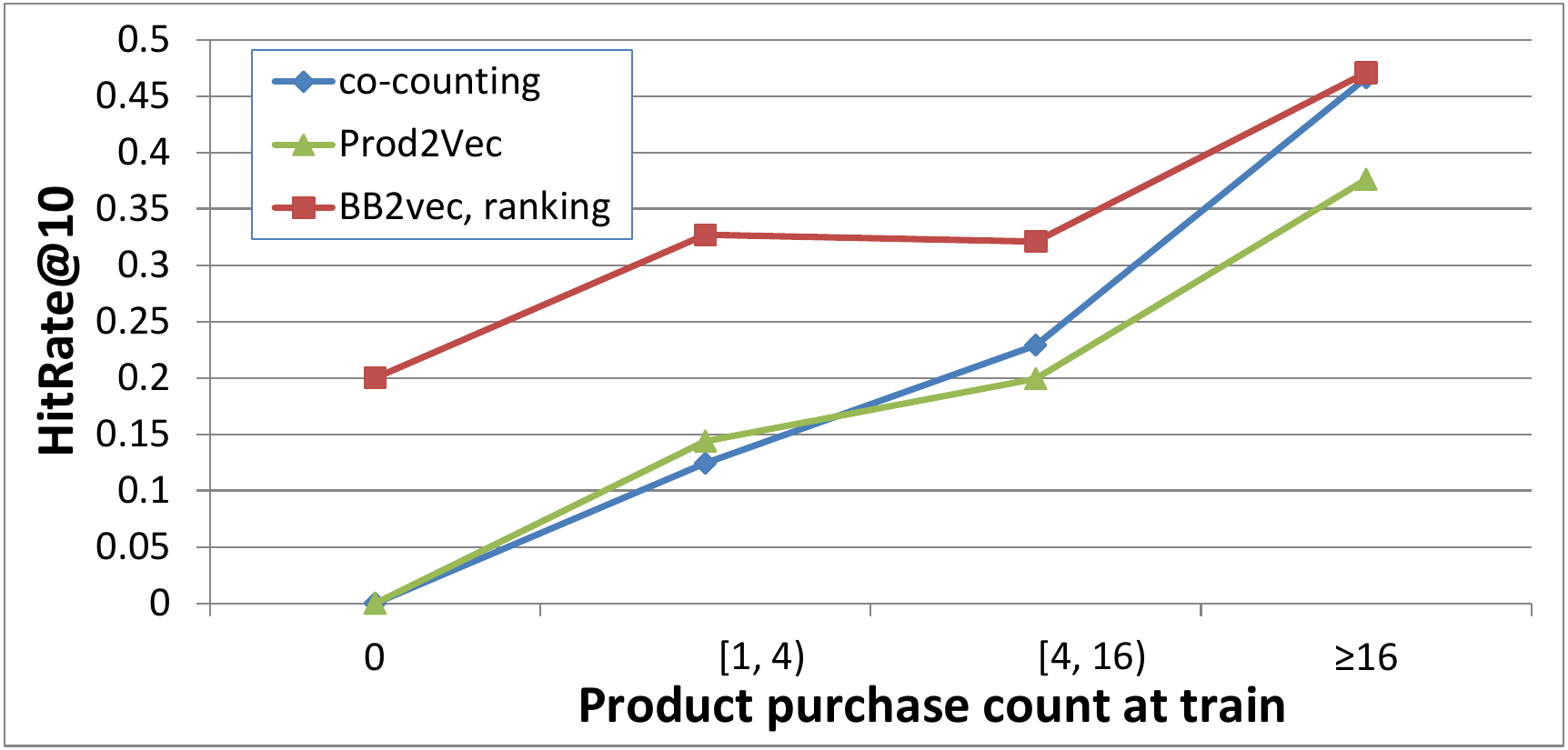}
                \caption{CIKM'16, Categories}
        \end{subfigure}
        \begin{subfigure}[tp]{0.49\textwidth}
                \includegraphics[width=\textwidth]{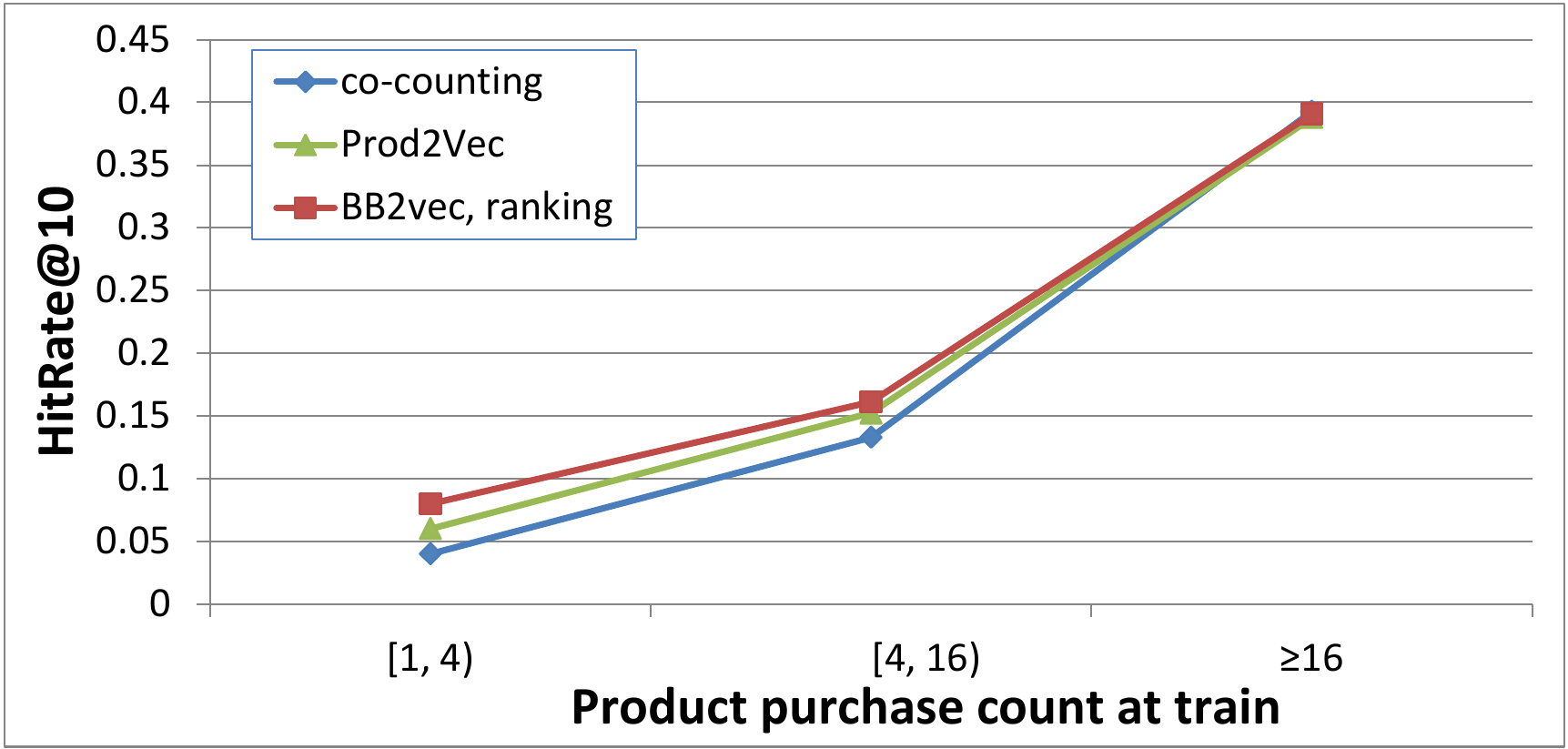}
                \caption{ACM RecSys'15}
        \end{subfigure}
        \begin{subfigure}[tp]{0.49\textwidth}
                \includegraphics[width=\textwidth]{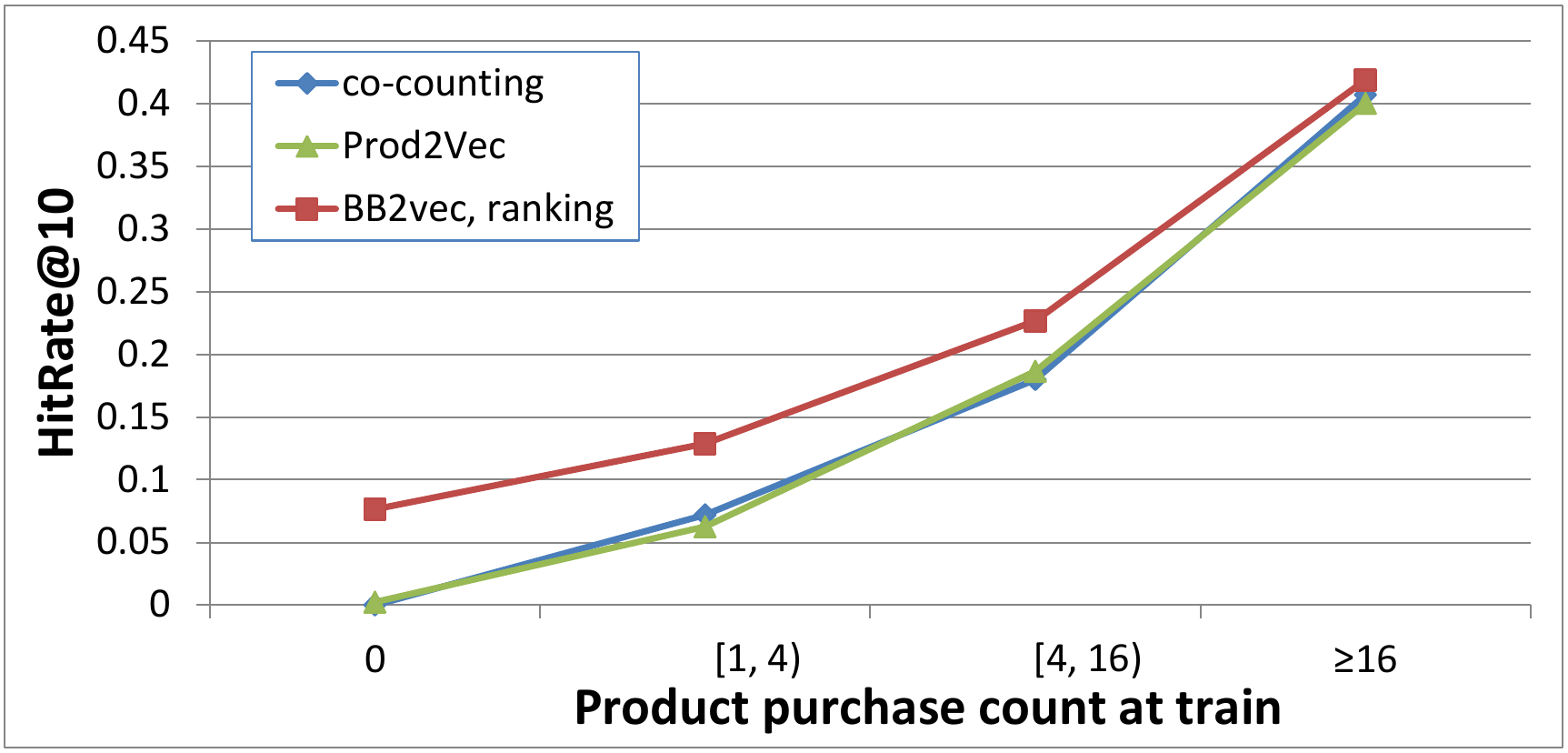}
                \caption{ACM RecSys'15, 10\%}
        \end{subfigure}
        \caption{Model performance vs. product purchase count}
        \label{fig:quality-vs-purchase}
\end{figure*}

In computational experiments, we evaluated the performance of the proposed \mymodel{} model.
This model learns vector representations of products from baskets and browsing sessions by optimizing the objective (\ref{obj-purchased-together-sg}).
We evaluated two variants: with classification (\mymodel{}\texttt{, class.}) and ranking (\mymodel{}\texttt{, ranking}) objectives.

We compared the proposed \mymodel{} model with the following baselines:
\begin{enumerate}
\item \texttt{Popularity}: items are sorted by purchases count at train.
\item \texttt{Co-counting}: for the query item $k$ top items with joint purchases $n_{km}$ are selected.
If for two items $m, r:$ $n_{mk} = n_{rk}$ then the item with larger purchase count had higher rank.
\item \texttt{prod2vec}. We implemented the \texttt{prod2vec} algorithm \citep{Grbovic2015} and fit item representations from basket data.
The context in the skip-gram objective always was all items in a basket except the input item.
It is equivalent to setting window size in \texttt{prod2vec} to the size of the largest basket. Thus the model optimizes the objective (\ref{prod2vec}).
\end{enumerate}

Validation datasets were used for early stopping.
The initial learning rate in AdaGrad was set to $0.05$. The number of negative samples was set to 20.
All hyperparameters of the methods (latent vectors dimensionality, mixing parameters $\lambda$) of were tuned for maximizing HitRate@10 at validation datasets and final results in the table \ref{tbl:res} are calculated at test datasets (see Appendix \ref{app:hyperparams}).
In the \texttt{prod2vec} and \mymodel{} models negative items were sampled uniformly.


For the predictive models \texttt{prod2vec} and \mymodel{} we used matrices $V'_B, V_B$ for generating recommendations.
Given a query item $k$, top N items by the score ${\vv'_{m, B}}^T \vv_{k, B} $ are selected.


\subsection{Results}

Firstly, we analyze the overall performance of the methods under evaluation.
Table \ref{tbl:res} show the average HitRate and NDCG over test datasets.
We conclude that the proposed \mymodel{} model has the best predictive performance among almost all the datasets and performance measures.
The difference between the models \mymodel{}\texttt{, class.} vs. \texttt{prod2vec} is the effect of learning of shared product representations
for both purchases and browsing sessions, instead of representations for purchases only.
The variant with the learning-to-rank objective \mymodel{}\texttt{, rank.} is better than the classification one \mymodel{}\texttt{, class.}
The gap between \mymodel{} and other models is the most sound at sparse datasets CIKM'16, CIKM'16 Categories, RecSys'15 10\%.

Secondly, we analyze in how the \mymodel{} model alleviates the cold start problem.
Fig. \ref{fig:quality-vs-purchase} shows the breakdown of average HitRate@10 by product purchase count at train dataset.
The \mymodel{} model performs better for products having few or no purchases.
In the same time, for products having enough purchases ($\ge 4$ for CIKM dataset, $\ge 16$ for other datasets) the \mymodel{} does not generate better recommendations than other models.
The notorious feature of the \mymodel{} model is that is can make recommendations for products with no purchases at the training set.
We conclude that the improvement of the overall performance comes from better recommendations for products having few or no purchases at the training dataset.

\section{Conclusions}

In this paper, we proposed the \mymodel{} model for complementary products recommendation in e-commerce.
The model learns product representations by processing simultaneously two kinds of data - baskets and browsing sessions
which are the very basic ones and always available to any e-commerce company.
Despite the large amount of browsing sessions, the learning algorithm is computationally scalable.

The predictive performance of the \mymodel{} model is better than the performance of state-of-the-art models which rely solely on basket data.
We show that the improvement comes from better recommendations for products having few or no purchases.
Thus, the proposed model alleviates the cold start problem.
Unlike other models, the \mymodel{} model can generate recommendations for products having no purchases.




Our model can be extended in two ways.
Firstly, one can use other algorithms as tasks in the multi-task objective, for example the \texttt{P-EMB} model.
Secondly, our model can be extended to handle items metadata.

\begin{acks}
The author thanks Konstantin Bauman and Pavel Serdyukov for their useful comments and proofreading.

\end{acks}
\appendix
\section{Ranking loss function}
\label{ranking-explained}

Consider a basket $B$, input item $k \in B$, output item $m \in B$ and negative samples $r \sim P^{\alpha}(w)$.
Then the negative sampling objective for learning-to-rank variant equals

\begin{align*}
Q^{SG}(V, V') &= \sum_{B \in \mathcal{B}} \sum_{\substack{k, m \in B \\ k \neq m}} \mathbb{E}_{r \sim P^{\alpha}(w)} \log (1 + \exp((\vv'_r - \vv'_m)^T \vv_k))
\end{align*}

Let $P(k, m)$ be the empirical distribution of item pairs $(k,m)$ in the baskets $\mathcal{B}$, $a_{rmk} = (\vv'_r - \vv'_m)^T \vv_k$,
$N =  \sum_{B \in \mathcal{B}} |\{ k, m \in B, k \neq m \}|$.
Then
\begin{align*}
\frac{1}{N} Q^{SG}(V, V') &= \sum_{k, m, r} P(k, m) P^{\alpha}(r) \log (1 + \exp((\vv'_r - \vv'_m)^T \vv_k)) \\
&= \sum_{k, m, r} P(k) P(m | k) P^{\alpha}(r) \log (1 + \exp((\vv'_r - \vv'_m)^T \vv_k)) \\
&= \sum_{k, m, r} P(k) P(m | k) P^{\alpha}(r) \log (1 + \exp(a_{rmk})) \\
&= \frac{1}{2} \sum_{k, m, r} P(k) P(m | k) P^{\alpha}(r) \log (1 + \exp(a_{rmk})) \\
 & \qquad + P(k) P(r | k) P^{\alpha}(m) \log (1 + \exp(-a_{rmk})) \\
&= \frac{1}{2} \sum_{k, m, r} P(k) f(a_{rmk})
\end{align*}
here we defined the function
$$
f(a) = P(m | k) P^{\alpha}(r) \log (1 + \exp(a)) + P(r | k) P^{\alpha}(m) \log (1 + \exp(-a))
$$
and used the identity $a_{mrk} = -a_{rmk}$.

The derivative of $f(a)$ is
\begin{align*}
f'(a) = \frac{1}{1 + \exp(a)} (P(m | k) P^{\alpha}(r) \exp(a) - P(r | k) P^{\alpha}(m))
\end{align*}
By solving the equation $f'(a_{rmk})=0$ we obtain
\begin{align*}
a_{rmk} = \log\left(\frac{P(r | k)}{P^{\alpha}(r)}\right) - \log\left(\frac{P(m | k)}{P^{\alpha}(m)}\right)
\end{align*}


Recalling than $a_{rmk} = (\vv'_r - \vv'_m)^T \vv_k$ we conclude that ranking by the score ${\vv'_r}^T \vv_k$
is equivalent to the ranking by $P(r | k) / P^{\alpha}(r)$.

\section{Best hyperparameters of the models}
\label{app:hyperparams}
  \label{tbl:hyperparams}
  \begin{tabular}{ccccccccc}
    \toprule
    Dataset \slash \, Model    & \texttt{prod2vec} &  \mymodel{} \\
    \midrule
    CIKM'16            & $d=400$ & $d=100, \lambda = 2$ \\
    CIKM'16 Categories & $d=200$ & $d=100, \lambda = 8$ \\
    RecSys'15          & $d=100$ & $d=200, \lambda = 8$ \\
    RecSys'15, 10\%    & $d=100$ & $d=400, \lambda = 32$ \\
  \bottomrule
\end{tabular}

\bibliographystyle{ACM-Reference-Format}
\bibliography{acc-rec-bibl}

\end{document}